\documentclass[summary]{gass2017}
\usepackage{color}
\usepackage{graphicx}
\title{CHIME FRB: An application of FFT beamforming for a radio telescope}
\author{Cherry Ng*\affref{ref1}, Keith Vanderlinde\affref{ref2}\affref{ref3},
  Adiv Paradise\affref{ref2}, Peter Klages\affref{ref2}, Kiyoshi Masui\affref{ref1}, 
  Kendrick Smith\affref{ref4}, Kevin~Bandura\affref{ref5}\affref{ref6}, 
  Patrick Joseph Boyle\affref{ref7}, Matt~Dobbs\affref{ref7}\affref{ref8},
  Victoria Kaspi\affref{ref7}\affref{ref8}, Andre Renard\affref{ref3}, J.~Richard~Shaw\affref{ref1}, 
  Ingrid~Stairs\affref{ref1}\affref{ref8}, Ian~Tretyakov\affref{ref3}}

\affiliation{%
  \aff{ref1}{Department of Physics and Astronomy, The University of British Columbia, Vancouver, BC, V6T-1Z1, Canada}
  \aff{ref2}{Department of Astronomy \& Astrophysics, University of Toronto, Toronto, ON, M5S-3H4, Canada}
  \aff{ref3}{Dunlap Institute for Astronomy \& Astrophysics, University of Toronto, Toronto, ON, M5S-3H4, Canada}
  \aff{ref4}{Perimeter Institute for Theoretical Physics, Waterloo, ON, N2L-2Y5, Canada}
  \aff{ref5}{LCSEE, West Virginia University, Morgantown, WV 26505, USA}
  \aff{ref6}{Center for Gravitational Waves and Cosmology, West Virginia University, Morgantown, WV 26505, USA}
  \aff{ref7}{Department of Physics \& McGill Space Institute, McGill University, Montreal, QC, H3A-2T8, Canada }
  \aff{ref8}{Canadian Institute for Advanced Research, Toronto, ON, M5G-1Z8, Canada }}

\begin{document}

\maketitle

\begin{abstract}
We have developed FFT beamforming techniques for the CHIME radio telescope, to search for and localize the astrophysical signals from Fast Radio Bursts (FRBs) over a large instantaneous field-of-view (FOV) while maintaining the full angular resolution of CHIME. We implement a hybrid beamforming pipeline in a GPU correlator, synthesizing 256 FFT-formed beams in the North-South direction by four formed beams along East-West via exact phasing, tiling a sky area of $\sim250$ square degrees. A zero-padding approximation is employed to improve chromatic beam alignment across the wide bandwidth of 400 to 800\,MHz. We up-channelize the data in order to achieve fine spectral resolution of $\Delta\nu=$24\,kHz and time cadence of 0.983\,ms, desirable for detecting transient and dispersed signals such as those from FRBs.
\end{abstract}

\section{Introduction}
The Canadian Hydrogen Intensity Mapping Experiment (CHIME) is a radio telescope currently being constructed at the Dominion Radio Astrophysical Observatory (DRAO) in Penticton, BC, Canada \cite{CHIME}. CHIME is a transit telescope with no moving parts, composed of four cylindrical reflecting surfaces, each 100\,m in length North-South (N-S) and 20\,m wide East-West (E-W). On each of the four focal lines is a linear array of 256 dual polarization antennas with a spacing of 0.3048\,m. This geometry provides an extremely wide effective FOV of $\sim$120$^{\circ}$ in N-S and 1.3$-$2.5$^{\circ}$ (frequency dependent) in E-W, that is, a primary beam size of $\sim250$ square degrees. CHIME is expected to be operational in 2017.
 
Apart from the original science goal of mapping redshifted 21-cm hydrogen emission to study the Baryon Acoustic Oscillation signal at redshifts 0.8$-$2.5 \cite{CHIME}, a custom backend is being deployed to conduct a commensal blind survey of FRBs. A recently discovered phenomenon \cite{Lorimer,Thornton}, FRBs appear to be a new class of radio transient with unknown astrophysical origin. The current challenges in the FRB field are mainly twofold: first, traditional radio telescopes employed for the study of point sources tend to have a limited FOV. This has meant low survey efficiency in the FRB search effort thus far, and the small sample of only twenty or so known FRBs has limited the precision with which FRB rates can be estimated. Secondly, only one FRB has been observed to repeat \cite{Spitler}. It remains the only FRB that has been localized by subsequent interferometric observations, which allowed for the identification of a host galaxy \cite{Tendulkar}. For all other FRBs, the localization is only approximately as good as the full width at half-maximum (FWHM) of the beam of the initial single dish detection, typically of the order of a few square arcminutes. This has made it impossible to identify multi-wavelength counterparts, which is crucial for determining the origin of FRBs.

Interferometers that are utilized to study time-domain astronomy are typically employed in a phased array mode, where targeted beams are formed by coherently summing antenna signals. Ref.~\cite{Tegmark2009,Tegmark2010} have shown that for an array of regularly spaced antennas as in CHIME, the interferometric phasing required for a complete basis of formed beams resembles the functional form of a Fourier Transform. FFT beamforming is computationally interesting for the context when the number of antennas $N$ is large, as it reduces the computational cost from $\mathcal{O}(N^{2})$ to $\mathcal{O}(N\log_{2}N)$. Commercially-available graphics processing units (GPUs) have now provided an affordable solution to the powerful computing necessary for this task. 

We have implemented a real-time FFT-beamforming pipeline (see Section~\ref{sec:pipeline}), which will generate an array of discrete but closely spaced, high-spectral resolution synthetic beams, allowing us to efficiently search the entire primary beam for short-duration transient signals. By searching these beams, CHIME has the potential to discover tens of FRBs per day \cite{Connor,Chawla2017}, and is well positioned to revolutionize the field of FRB research.

\section{CHIME Correlator} \label{sec:Xengine}
The CHIME correlator follows a hybrid FX design. Input voltage data are collected from 256 dual-polarization receivers on each of the 4 cylinders, i.e., a total of 2048 distinct inputs. Custom FPGA boards (F-engine) \cite{ICE1,ICE2} digitize and channelize these data to 1024 frequency channels via a 4-tap polyphase filter bank (PFB), then scale this voltage data to 4+4-bit complex numbers, at a 2.56\,$\mu$s cadence (see Table~\ref{tab:table}). Spatial correlation (X-engine) is performed in a GPU cluster that consists of 256 processing nodes each with 4 AMD Fiji GPUs, similar to that presented in \cite{Recnik,Denman,Klages}. This GPU cluster is also where the FRB fan-beam pipeline takes place.

\begin{table}
    \centering
  \caption{The specifications of the input and output data of the CHIME/FRB beamformer.}
 \begin{minipage}{9cm}
\begin{tabular}{p{4.0cm}p{1.6cm}p{1.8cm}}
\hline
Parameter & Input & Output\\
          & baseband       & intensity \\
\hline
Time resolution & 2.56\,$\mu$s & 0.983\,ms\\
Frequency resolution & 390\,kHz & 24\,kHz\\
Number of spectral channels & 1024 & 16$\times$1024 \\
Number of bits & 4-bit real + 4-bit imag & 8-bit integer \\
Number of polarizations & 2 & 1 \\
Total data rate & 6.4\,Tbps & 131.1\,Gbps\\
\hline \label{tab:table}
 \end{tabular}
\vspace*{-9mm}
\end{minipage}
\end{table}

\section{Implementation in the context of CHIME/FRB} \label{sec:pipeline}
The FRB fan-beam pipeline has been implemented as three OpenCL kernels. The first kernel carries out a hybrid beamforming task. In the N-S direction, FFT beamforming is used to form 256 beams. In the E-W direction, exact phasing is employed to form four beams. The second kernel transposes the data from time--polarization--beams$_{\rm{EW}}$--beams$_{\rm{NS}}$, where time is the slowest varying index, to the order of polarization--beams$_{\rm{EW}}$--beams$_{\rm{NS}}$--time, where time becomes the fastest varying index as required by the next step. A third kernel applies a fine channelization, increasing spectral resolution by 128 times to 3\,kHz channels.

As a final step, Stokes I beams are formed by squaring the real and imaginary parts. The output data are integrated to 16k frequency channels (i.e., at spectral resolution $\Delta\nu=24\,$kHz) at 0.983\,ms cadence and rescaled to 8-bit integers (see Table~\ref{tab:table}). A total output data rate of 131.072\,Gbps, or 0.512\,Gbps per node, is transmitted via 1\,GbE output data links in a custom UDP packet format to the subsequent FRB search, where dedispersion and triggering take place. The current implementation of the end-to-end fan-beam pipeline utilizes 15\% of the available GPU computational budget, within the planned allocation among other tasks of the correlator GPU cluster. 

In the following we discuss some of the key elements of the FRB fan-beam pipeline in further detail:

\subsection{Fan Beamforming}
\begin{figure}[htbp]
  \centering
  \includegraphics[width=80mm]{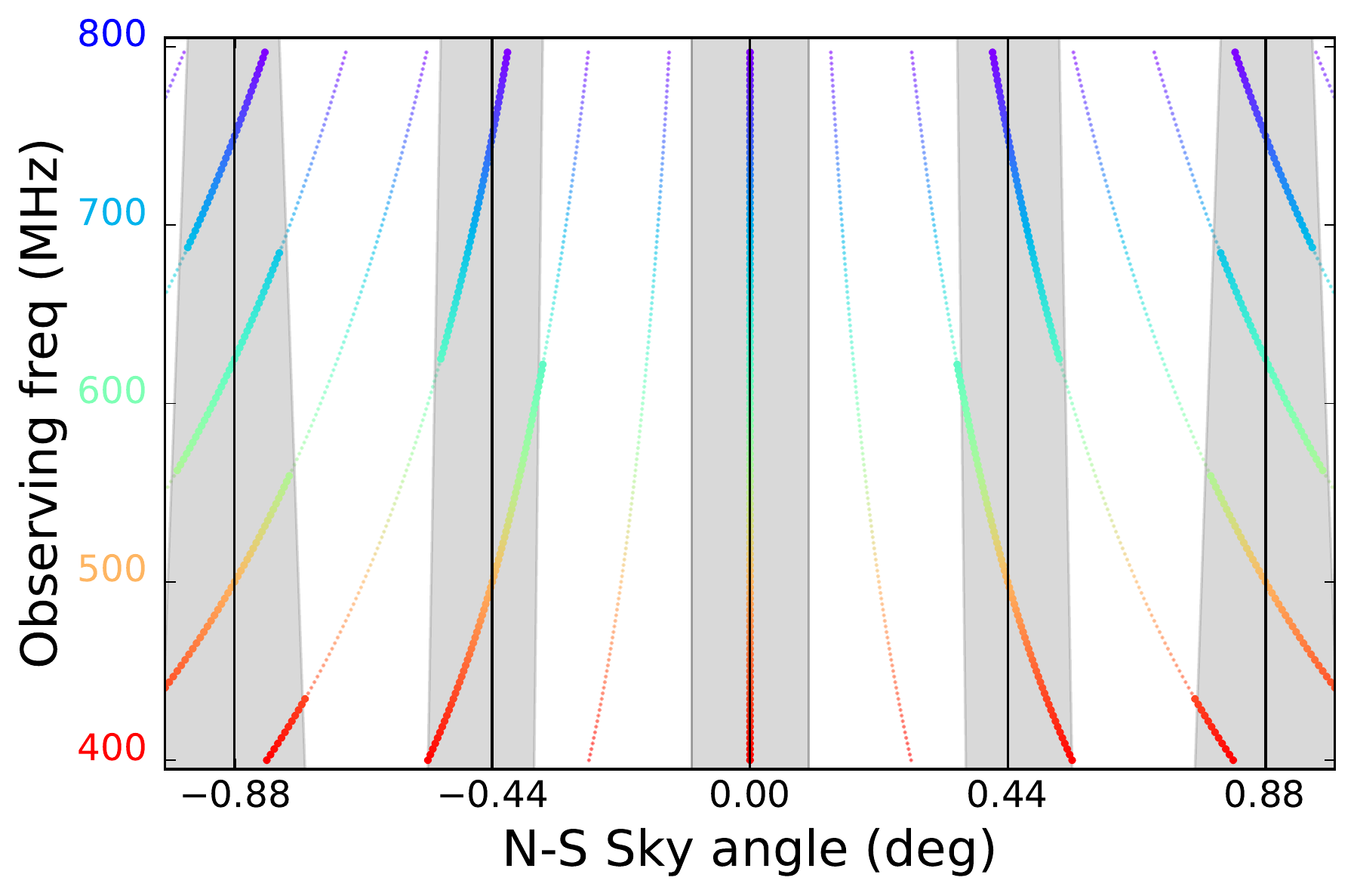}
  \caption{The chromatic effect of FFT beamforming across a wide bandwidth. The position of the dots represents the nominal peak sensitivity of each formed beam at a specific frequency, and different colours represent different observing frequencies. The final extent of the frequency-summed formed beams are illustrated by the gray areas.}
  \label{fig:chromatic}
\end{figure}

In the N-S direction, CHIME has $N=256$ antennas in a linear and regular array, which makes spatial FFT beamforming a sensible choice. Figure~\ref{fig:chromatic} shows that the nominal position of peak sensitivity of the FFT-formed beams is a function of the observing frequency $f_{\rm{k}}$. The formed beams at higher frequencies are closer together than those at lower frequencies. Hence, for a wide bandwidth instrument like CHIME, a straightforward implementation of FFT beamforming will lead to chromatic smearing, where the formed beams across the 400\,MHz band line up poorly.

Instead, we employ an approximate method using a zero-padded FFT. By extending the input data with a large number of zeros before taking the FFT, additional redundant beams can be formed. The output samples are then sufficiently dense that there will be a sample close to a given desired steering angle. The more we increase the padding, the closer one can get to any arbitrary steering angle. However, the computational cost also increases to roughly $\mathcal{O}(NP\log_{2}(NP))$, where $P$ is the padding factor. Taking both the computational cost and the beam sensitivity into account, a padding factor of $P=$2 has been chosen to create 512 beams, which will be subsampled by selecting the subset of beams at each frequency closest to the 256 desired pointings. In Figure~\ref{fig:chromatic}, the coloured dots enclosed by the gray areas are the selected nearest-neighbour clamped beams, whereas the fainter coloured dots are the additional and discarded beams.

\begin{figure*}[!htbp]
\centering
\setlength\fboxsep{0pt}
\setlength\fboxrule{0pt}
\fbox{\includegraphics[width=15.5cm]{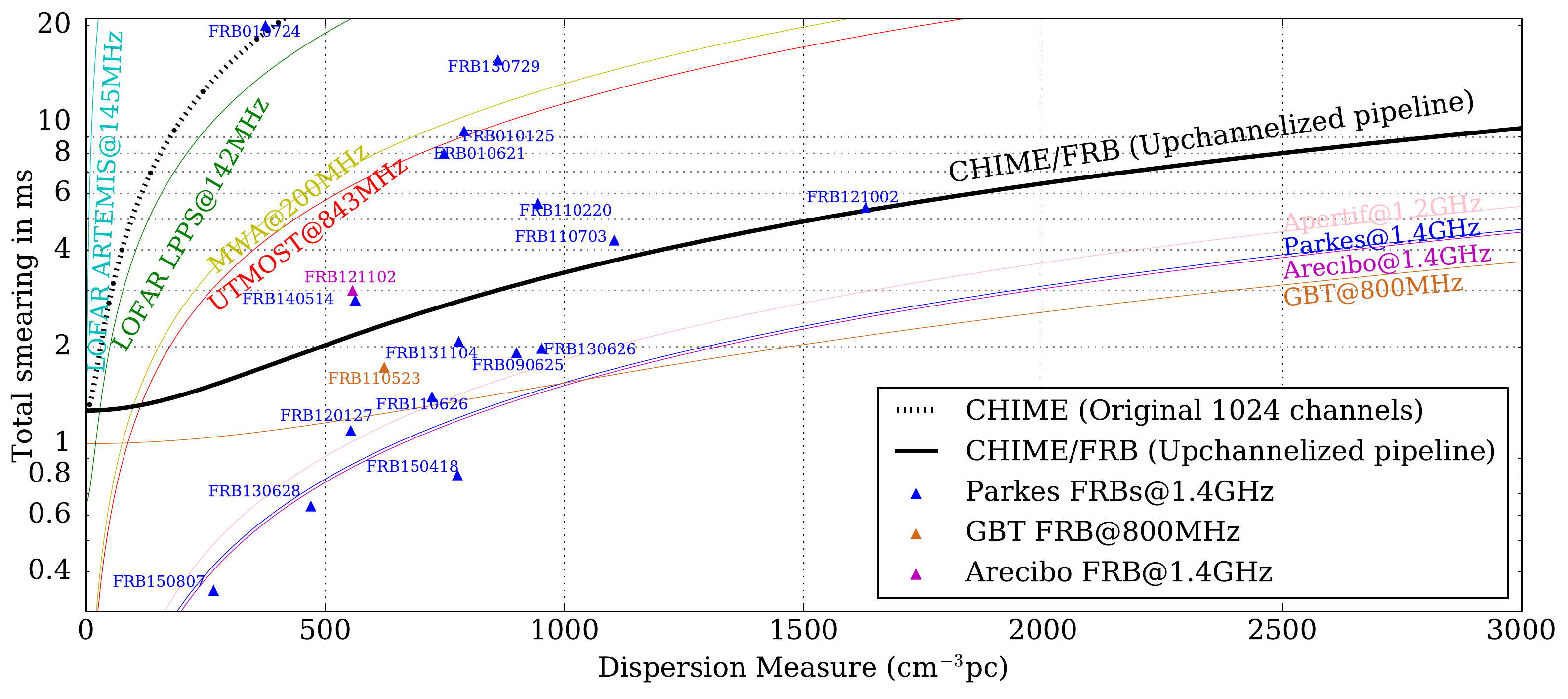}}
\caption{Comparing the smearing timescale vs search dispersion measure for different pre-search data treatment. The black dotted line is the original CHIME input of 1024 frequency channels. The current up-channelized pipeline is shown by the solid black line. For reference, the L-band observed widths of all known FRBs are also shown. These represent lower limits for CHIME, as interstellar broadening will most likely increase signal width at the lower 400$-$800\,MHz observing frequency of CHIME. The coloured lines are the intra-channel smearing of some examples of other FRB surveys. Apart from Parkes (blue), Arecibo (magenta, almost identical to the blue) and GBT (brown) which have discovered FRBs, there are two on-going surveys at LOFAR (cyan and green), one with MWA (yellow), one with Apertif (pink) and one with Molonglo (red). }
  \label{fig:smearing}
\end{figure*}

As shown in \cite{Maranda}, the clamping can be efficiently done by calculating a set of reference beam indices ($t_{\rm{m}}(k)$) for a reference frequency $f_{\rm{k}}$ using Equation~(1), and then finding the adjustment to those indices ($\Delta t_{\rm{m}}$) at each observing frequencies ($\Delta f$) using Equation~(2). In both equations, the speed of light is denoted by $c$ and $d$ is the antenna separation. These indices can be pre-computed and stored as look-up tables for the beamforming pipeline.
\begin{equation}
t_{m}(k) = N \times P \times f_{k} \times \frac{d}{c} \sin\theta_{m,k} +0.5\,.
\end{equation}
\begin{equation}
\Delta t_{m} = N \times P \times \Delta f \times \frac{d}{c} \sin\theta_{m,k}\,.
\end{equation}

The frequency-summed beam pattern becomes progressively more elongated towards the horizon, with FWHM of $\sim0.5^{\circ}$ at zenith and up to $\sim2.8^{\circ}$ at $10^{\circ}$ elevation. The extent of the N-S horizon coverage is tunable, and we have the flexibility to form denser or sparser beam tilings. Also note that the E-W formed beams can be configured as static or tracking.

\subsection{Up-channelization}
The interstellar medium acts as a cold, ionized plasma, dispersing the broadband FRB emission as it propagates towards the observer. Dispersion across frequency channels can be corrected for by delaying the time series at higher frequencies by the appropriate amount to align with the signal arrival time at the lower frequencies. Intra-channel dispersion, however, can only be corrected for through coherent de-dispersion, which is computationally too expensive in a search operation. If not accounted for, intra-channel dispersion will smear the observed pulse-like FRB signal, limiting our sensitivity to FRBs. The beamformer input has only 1024 frequency channels across the 400-MHz band, i.e., a spectral resolution of 390\,kHz. As shown in Figure~\ref{fig:smearing}, this is far from ideal compare to the pulse widths of the known FRBs. The FRB fan-beam processing, therefore, includes a second stage channelization, transforming every 128 time samples with an FFT, to achieve a 3\,kHz spectral resolution. The improved frequency resolution is crucial for detecting narrow-width transient signals, hence maximizing our sensitivity towards FRB events. We are also investigating the feasibility of including a fourth kernel which will incoherently dedisperse the post-upchannelized data to a high nominal DM in order to boost our fiducial sensitivity to high-DM FRB events.

\subsection{Aliasing}
\begin{figure*}[!htbp]
\centering
\setlength\fboxsep{0pt}
\setlength\fboxrule{0pt}
\fbox{\includegraphics[width=16cm]{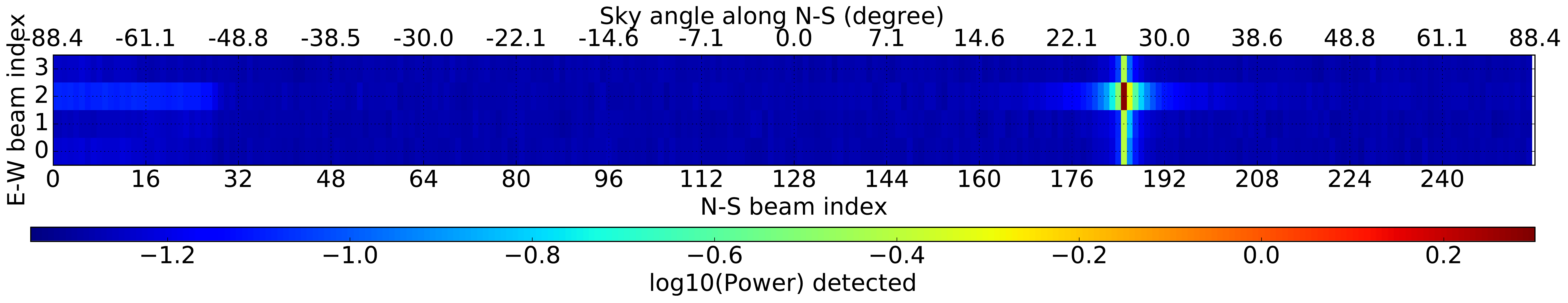}}
  \caption{Simulated data with an injected source at a given sky location. This plot shows one time stamp with all 16k frequency channels summed. The source is successfully re-detected by our GPU-based fan-beam pipeline, well localized with maximum power detected at the expected position on the NS-EW plane of the CHIME FRB field-of-view. }
  \label{fig:sim}
\end{figure*}

Spatial aliasing occurs when the physical spacing of the detectors equals half of the observing wavelength or shorter. CHIME has a N-S antenna spacing of 0.3048\,m, which means above an observing frequency of $\sim$490\,MHz we will inevitably have aliased signals. The strength of an aliased signal could be of comparable magnitude to the actual source in a single frequency channel. However, the location of the aliased beams are highly frequency dependent and hence in the frequency-summed response would appear `smeared' across a range of sky location relative to the main beam. The higher the zenith angle, the wider the extent of smearing in the corresponding aliased beams. Figure~\ref{fig:sim} is the frequency-summed sky response of simulated CHIME data. An injected signal is detected at the correct location near $\sim26^{\circ}$ from zenith, whereas its aliased signals are smeared across roughly $-88^{\circ}$ to $-50^{\circ}$. This plot also confirms that the aliased artifact appears much weaker compared to the unaliased beam in the frequency-summed response data, which is what downstream FRB search pipeline will be manipulating.

\section{Conclusion}
We have implemented a custom beamforming algorithm to the CHIME/FRB backend in order to generate 1024 intensity beams at high spectral and time resolution. This will enable 24/7 searching in CHIME's entire FOV and potentially bring about the discovery of a large population of fast transients. Work is on-going to compare simulated data with response power from the fan-beam pipeline. Figure~\ref{fig:sim} shows one example of an injected signal re-detected from the GPU pipeline output. Further proposed extensions of the CHIME/FRB beamforming project include beam interpolation, which would enable the FFT-formed beams to be perfectly aligned across observing frequencies (Masui et al., in prep). 

\section{Acknowledgements}
We are very grateful for the warm reception and skillful help we have received from the staff of the Dominion Radio Astrophysical Observatory, which is operated by the National Research Council of Canada. We acknowledge support from the Canada Foundation for Innovation, the Natural Sciences and Engineering Research Council of Canada, the B.C. Knowledge Development Fund, le Cofinancement gouvernement du Quebec-FCI, the Ontario Research Fund, the CIfAR Cosmology and Gravity program, the Canada Research Chairs program, and the National Research Council of Canada. We thank Xilinx University Programs for their generous support of the CHIME project, and AMD for donation of test units. CN acknowledges support from the Dunlap Institute for Astronomy \& Astrophysics at the University of Toronto. We thank Mandana Amiri, Nolan Denman, Mark Halpern, Tom Landecker, Jeff Peterson, Ziggy Pleunis, Masoud Ravandi and Kris Sigurdson for useful comments.

\end{document}